\documentclass{aa}
\usepackage{graphicx}
\usepackage{longtable}
\usepackage{lscape}

\input{psfig.tex}
\def\gsim{\;\lower.6ex\hbox{$\sim$}\kern-7.75pt\raise.65ex\hbox{$>$}\;}
\def\lsim{\;\lower.6ex\hbox{$\sim$}\kern-7.75pt\raise.65ex\hbox{$<$}\;}

\begin{document}
\title{Na-O Anticorrelation And HB I. The Na-O
anticorrelation in NGC 2808
\thanks{
Based on observations collected at ESO VLT-UT2 under programme 72.D-0507}
 }

\author{
E. Carretta\inst{1},
A. Bragaglia\inst{1},
R.G. Gratton\inst{2},
F. Leone\inst{3},
A. Recio-Blanco\inst{4}
\and
S. Lucatello\inst{2}}

\authorrunning{E. Carretta et al.}
\titlerunning{Na-O anticorrelation in NGC 2808}

\offprints{E. Carretta, eugenio.carretta@bo.astro.it}

\institute{
INAF - Osservatorio Astronomico di Bologna, Via Ranzani 1, I-40127
 Bologna, Italy
\and
INAF - Osservatorio Astronomico di Padova, Vicolo dell'Osservatorio 5, I-35122
 Padova, Italy
\and
INAF - Osservatorio Astrofisico di Catania, Via S. Sofia 78, I-95123, 
Catania, Italy
\and
Dpt. Cassiop\'ee, UMR 6202, Observatoire de la C\^ ote d'Azur, B.P. 4229, 06304 
Nice Cedex 04, France 
  }

\date{ }

\abstract{
We derived atmospheric parameters and elemental abundances of Fe, O
and Na for about 120 red giant stars in the Galactic globular cluster NGC~2808.
Our results are based on the analysis of medium-high resolution (R=22000-24000)
GIRAFFE 
spectra acquired with the FLAMES spectrograph at VLT-UT2 as a part of a project
aimed at studying the Na-O anticorrelation as a function of physical parameters in
globular clusters. We present here the anticorrelation of Na and O abundances
in NGC~2808, and we discuss the distribution function of stars along this
relation. Besides a bulk of O-normal stars, with composition typical of field
halo stars, NGC~2808 seems to host two other groups of O-poor and super O-poor
stars. In this regard, NGC~2808 is similar to M~13, the template cluster for
the Na-O anticorrelation. However, at variance with M~13, most stars in 
NGC~2808 are O-rich. This might be related to the horizontal branch morphologies 
which are very different in these two clusters. 
The average metallicity we found for NGC~2808 is [Fe/H]$=-1.10$
(rms=0.065 dex, from 123 stars). We also found some evidence of a small 
intrinsic spread in metallicity, but more definitive conclusions are hampered
by the presence of a small differential reddening.
 }

\maketitle

\section{Introduction}

This is the first paper in a series aimed at uncovering and studying the
possible  existence of a second generation of stars in Galactic Globular
Clusters (GCs). The presence and the properties of these stars, likely born
from the ejecta of intermediate mass stars, will be inferred from the analysis
of the Na-O anticorrelation, found and extensively studied in a number of GCs
mainly by the Lick-Texas group (Kraft, Sneden and coworkers; see Gratton, Sneden
\& Carretta  2004 for a recent review and a summary of abundance variations in
clusters).

The Lick-Texas group found that in
most of the surveyed GCs there is a star-to-star anticorrelation between the O
and Na abundances. This is a sign of the (unexpected) presence of material processed
through the complete CNO cycle in GC stars: at the temperature where this
occurs, $^{22}$Ne is transformed into $^{23}$Na by proton capture (Denisenkov
\& Denisenkova 1989; Langer et al. 1993). Hence, enhanced Na abundances should
accompany O depletions in stars.
While early interpretations called for deep mixing processes in the same stars
where abundance anomalies are observed, Gratton et al. (2001) and subsequently
Carretta et al. (2004b) showed that the  CNO cycle processed material must be
due to pollution from ejecta of other (more massive) stars, since the Na-O 
anticorrelation is found also among unevolved stars in clusters of any 
metallicity (NGC 6397, NGC 6752, 47 Tuc were studied). 

Favorite nucleosynthesis sites are thermally pulsating intermediate-mass  
Asymptotic Giant Branch (AGB)
stars undergoing hot bottom burning (Ventura et al. 2001). It is  unlikely that
the Na-rich, O-poor material was acquired by the stars after their formation
(Cohen et al. 2002), because the accreted surface layers would be washed out 
by the deepening of the convective envelope during the red giant evolutionary
phase. Hence,  this anticorrelation most probably calls for a second generation
of stars, formed within GCs from the kinematically cool ejecta of massive AGB
stars (Cottrell and Da Costa 1981). 

The age difference between the two populations (a few $10^8$~yr) is too small
to be directly detectable as different Turn-Offs (TO's) in the color-magnitude
diagrams. However, we might expect a connection between the distribution of
stars along the Na-O anticorrelation and the - so far - unexplained presence
of extended Blue Horizontal Branches (BHBs) in several GCs. In fact, O-poor,
Na-rich (i.e., polluted) stars should also be enriched in He  (by about
$\Delta$Y=0.04: D'Antona et al. 2002). In turn, He-rich stars evolve faster on
the Main Sequence (MS), so that polluted stars currently at the TO should be
less massive (by about 0.05~$M_\odot$) than the "normal" He-poorer stars. If
these stars lost mass at the same rate as normal stars on the Red Giant Branch
(RGB), their descendants should become much hotter HB stars, maybe explaining
the long blue tails observed in many GCs (e.g. in M 13 and NGC 6752, which 
display the most extended Na-O anticorrelations).

However, this connection still needs to be statistically proven with 
significative samples of stars in different clusters, in order to disentangle
the possible link(s) between HB morphology (extension and mass distribution) and
the other parameters (like metallicity, age and the Na-O
anticorrelation distribution function, and even the close binary fraction).
A first attempt was made by Carretta et al. (2003) who used data from the 
FLAMES Science Verification program in order to study the
Na distribution along the RGB in NGC 2808, a cluster showing a well known
bimodal HB morphology, with a clump of red HB stars and a long distribution of
stars on the blue side of the RR Lyrae instability strip, down to very faint
magnitudes. FLAMES multiplex capability was used to
derive Na abundances for 81 RGB stars. 
Unfortunately, observations of O indicators were available only for a small
fraction of stars with measured Na abundances, since spectra
were acquired for another purpose. While
we are quite confident about the $shape$ of the anticorrelation, the 
Na or O abundances alone would be not enough to reconstruct the {\it
distribution function} of the anticorrelation, since Na or O saturate at the
edges of the distribution. On the other hand, the $ratio$ Na/O does continue to
vary even at extreme values along the anticorrelation.

To study the connection between the O-Na anticorrelation and the  HB morphology
an adequate sample of stars in each cluster is required: assuming a flat 
distribution (a very rough approximation), the probability $p$\ to cover at
least a fraction $x$\ of the total range using $n$\ stars is $p=1-x^n$. Hence,
to estimate the extent of the Na-O anticorrelation with 4\% accuracy at a 95\%
level of confidence we need to observe $\sim 80$\ stars in each GC.
About 20 GCs with a wide distribution of HB morphologies are required to
confidently conclude that a connection indeed exists. To estimate the full
extent of the Na-O anticorrelation, observations of stars down to [O/Fe]$\sim
-1$ dex\footnote{We use the usual  spectroscopic notation: log~n(A) is the
abundance (by  number) of the element A in the usual scale where log~n(H)=12;
notation [A/H] is  the logarithmic ratio of the abundances of elements A and H
in the star, minus  the same quantity in the Sun.}  are needed: this implies
high resolution, high S/N observations of RGB stars.

The capabilities of VLT+FLAMES (high multiplex gain, high resolution) allow us
to gather the required number  statistics, both in the number of GCs and of
stars studied in each GC. Hence, we started the present project in
order to perform a systematic analysis of large number of stars with accurate 
and homogeneous Na and O abundances in about 20 GCs.

In the present paper we present the method of analysis and the results 
obtained for NGC 2808. An outline of the observations is given in the next
Section. The derivation of atmospheric  parameters and the analysis are
discussed in Sect. 3, whereas error estimates are given in Sect. 4. Finally,
Sect. 5 and 6 are devoted to the reddening and  intrinsic scatter in Fe for NGC
2808 and to the results for the Na-O anticorrelation,  respectively. Summary
and conclusions are presented in Sect. 7.

\section{Observations}

Our data were collected with the ESO high resolution multifibre spectrograph 
FLAMES/GIRAFFE (Pasquini et al. 2002), mounted on VLT UT2. Observations were
done with two GIRAFFE setups, the high-resolution gratings HR11 (centered at
5728~\AA)  and HR13 (centered at 6273~\AA) to measure the Na doublets at
5682-5688~\AA\ and 6154-6160~\AA\ and the [O I] forbidden lines at 6300, 
6363~\AA, respectively. Resolution is R=24200 (for HR11) and R=22500 
(for HR13), at the centre of spectra.

\begin{table}
\caption{Log of the observations for NGC 2808. Date is UT, and exposure
times are in seconds}
\begin{tabular}{rccc}
\hline
Grating &Date       &UT$_{beginning}$ &exptime \\

\hline
HR11   & 2004-02-16  & 05:35:52.011  & 3000 \\ 
       & 2004-02-16  & 06:26:45.660  & 3000 \\ 
       & 2004-02-17  & 05:23:11.985  & 2850 \\ 
       & 2004-02-17  & 06:11:42.443  & 2850 \\ 
HR13   & 2004-01-12  & 05:36:55.750  & 2850 \\ 
       & 2004-01-12  & 06:25:25.730  & 2850 \\ 
       & 2004-01-18  & 05:56:41.205  & 3000 \\ 
       & 2004-01-18  & 06:47:34.701  & 3000 \\ 
\hline
\end{tabular}
\end{table}

We obtained 4 exposures for each grating (for both gratings, 2 exposures of
3000 s, and 2 of 2850 s); the observation log is given in Table 1.  Pointings
were centered at RA = 09:12:00.1, DEC =  $-$64:51:50.7  (J2000).

Stars were selected from the photometry by Bedin et al. (2000), kindly made
available by the authors, and already used for a FLAMES Science Verification
program (Carretta et al. 2003, 2004a). We chose stars along the RGB, from about
$V=13.9$ (1 mag below the RGB tip) down to $V=15.5$. Targets were selected among
isolated stars, i.e. all stars were chosen to be free from any companion
closer than 2 arcsec and  brighter than $V+2$ mag, where $V$ is the target
magnitude.

\begin{figure}
\begin{center}
\includegraphics[bb=60 190 380 680, clip, scale=0.6]{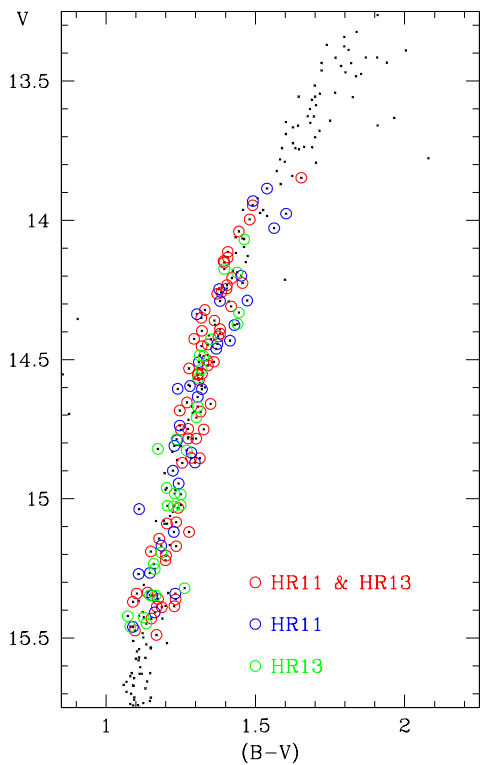}
\caption{CMD of NGC 2808 (taken from Bedin et al. 2000); 
the targets observed with the GIRAFFE/MEDUSA gratings are
indicated by open circles. The color coding indicates the setups used in the
observations (red for both HR11 and HR13, blue for HR11 only, green for HR13 
only).}
\label{f:figcmd}
\end{center}
\end{figure}

We decided to target different objects with the UVES fibers,
in order to observe up to 14 stars per
cluster in the highest resolution mode\footnote{The analysis of these higher
resolution spectra will be presented elsewhere}; hence, the GIRAFFE fiber 
positioning 
was also slightly different between the pointings used with the two gratings.
As a consequence, not all the stars were observed in both gratings; on a grand
total of 130 different stars observed, we have 65 objects with spectra for both
gratings, 34  with only HR11 observations and 31 with only HR13
observations. 
Since the Na doublet at 6154-6160~\AA\ falls into the spectral range covered by
HR13, we could measure Na abundances for all 130 target stars, whereas we could
expect to measure O abundances only  up to a maximum of 65+31 stars. A list of
all observed targets is given in Table~\ref{t:coo}, and the  colour magnitude
diagram (CMD) is shown in Figure~\ref{f:figcmd}.

\begin{table*}
\caption{List and relevant informations for the target stars observed in NGC
2808. ID, $B$, $V$ and coordinates (J2000) are taken from Bedin et al. (2000);
$J$, $K$ are from the 2MASS catalog; radial velocities RV's (in km s$^{-1}$)
from both gratings are heliocentric; stars with '*' in notes have $V-K$ colours
that deviate from the ones expected for RGB stars (see text). The complete
Table is available electronically; we show here a few lines for guidance.}
\begin{tabular}{rllccccrrrr}
\hline
 Star &RA (h m s)    & DEC (d p s)   &  V     &   B    &   J    &  K      &RV(HR11) &RV(HR13)& HR    &Notes\\
\hline
 7183 & 9 12  2.8710 & -64 49 34.069 & 14.854 & 16.168 &    	&         & 106.58 & 106.55  & 11,13 &  	\\ 
 7315 & 9 11 58.581  & -64 49 29.88  & 14.683 & 15.930 & 12.255 &  11.425 &  98.30 &  98.97  & 11,13 &  	\\ 
 7536 & 9 12 31.7065 & -64 49 22.268 & 14.372 & 15.812 & 11.802 &  10.829 &        &  96.77  & 13    &  	\\ 
 7558 & 9 12 20.1287 & -64 49 21.891 & 15.389 & 16.576 & 13.160 &  12.407 & 117.01 & 118.12  & 11,13 &  	\\ 
 7788 & 9 11 57.1979 & -64 49 14.551 & 14.870 & 16.168 & 12.472 &  11.623 &  98.91 & 	     & 11    &  	\\ 
 8198 & 9 11 48.9714 & -64 48 59.526 & 15.340 & 16.572 & 13.042 &  12.277 & 100.09 & 	     & 11    &  	\\ 
 8204 & 9 11 58.577  & -64 48 59.27  & 15.119 & 16.397 & 12.793 &  11.879 & 109.67 & 109.95  & 11,13 &  	\\ 
 8603 & 9 12 14.0510 & -64 48 42.915 & 14.432 & 15.847 & 11.902 &  10.957 & 109.93 & 	     & 11    &  	\\ 
 8679 & 9 11 44.5585 & -64 48 39.890 & 14.961 & 16.164 & 12.753 &  12.022 &        &  43.27  & 13    &  Field	\\ 
\hline
\end{tabular}
\label{t:coo}
\end{table*}

\section{Atmospheric parameters and analysis}

\subsection{Atmospheric parameters}

Temperatures and gravities were derived as described in
Carretta et al. (2003); along with the derived atmospheric parameters and
iron abundances, they are shown in Table \ref{t:atmpar} (completely available
only in electronic form). We used $K$ magnitudes taken from the Point
Source Catalogue of 2MASS (Cutri et al. 2003); the 2MASS photometry was
transformed to the TCS photometric system, as used in Alonso et al. (1999).

We obtained T$_{\rm eff}$'s and bolometric corrections B.C. for our stars
from $V-K$ colors whenever possible. 
We employed the relations by Alonso et al. (1999, eqs. 4, 7 and 17, with the
erratum of 2001). We adopted for NGC 2808 a distance modulus of $(m-M)_V$=15.59
and a reddening of $E(B-V)$ = 0.22 (Harris 1996, updated at
{\tt http://physun.physics.mcmaster.ca/Globular.html}), and the relations 
$E(V-K) = 2.75 E(B-V)$, $A_V = 3.1 E(B-V)$, and $A_K = 0.353 E(B-V)$ 
(Cardelli et al. 1989). 
An input metallicity  of [Fe/H]$=-1.14$ was adopted from Carretta et al.
(2004a), based on the analysis of UVES Red Arm spectra\footnote{This  value is
slightly different from what we derive in the present study;  however, the
dependence of ($V-K$) on [Fe/H] is so weak that  temperatures are almost
unaffected by this difference}. 

There are a few target stars with no $K$ magnitude from 2MASS and a few others
appearing as outliers in the $V$, $V-K$ diagram, but not in the $V$, $B-V$
diagram: this is probably due to the worse spatial resolution of 2MASS, an
important factor in a dense GC field. Temperatures for these stars were derived from a mean relation 
T$_{\rm eff}(V-K)$ as a function of T$_{\rm eff}(B-V)$.

Surface gravities log $g$'s were obtained from effective temperatures and 
bolometric corrections,   assuming that the stars have masses of 0.85
M$_\odot$. The adopted  bolometric magnitude of the Sun is $M_{\rm Bol,\odot} =
4.75$. 

We did not take into account the existence of differential reddening (Walker
1999, Bedin et al. 2000) since no individual correction  for each star is
available in literature. The differential reddening is however small:
peak-to-peak differences amount to $\sim 0.08$ mag ($\sigma \simeq 0.020 -
0.028$, Walker 1999, Bedin et al. 2000). The effect of this will be taken into
account as an additional error source in the abundance derivation (see Sect. 4).

\subsection{Equivalent widths}

Data reduction was done using standard IRAF\footnote{IRAF is
distributed by the National Optical Astronomical Observatory, which  are
operated by the Association of Universities for Research in Astronomy, under 
contract with the National Science Foundation} packages for bias subtraction,
flat-fielding correction, correction for scattered light, 
spectra extraction and wavelength calibration.
We measured radial velocities (RVs) for each spectrum (using RVIDLINES on about
25 to 40 lines for HR11 and HR13 respectively); the error is less than 0.5 km
s$^{-1}$ on each measure, and we put in Table~\ref{t:coo} the average for the 4
pointings in each grating, after correction for heliocentric motion.  The
average heliocentric velocity is RV = 102.4 km s$^{-1}$  ($\sigma= 9.8$), from
124 stars, after eliminating the non-members on the basis of their very
discrepant RVs (more than 6$\sigma$ from the average).

All spectra were shifted to zero radial velocities, then combined star by star;
this enhanced S/N and eliminated cosmic rays hits from the coadded spectra.

In the case of HR13, before coadding, each individual spectrum  was corrected
for blending  with telluric lines due in particular to H$_2$O and O$_2$ near
the [O I] line at 6300~\AA\ (we checked that no correction was necessary for the
[O I] line at 6363~\AA, or the Na lines in HR11).  We generated a synthetic
spectrum covering the interval 6280 to 6325~\AA, taking line positions and
equivalent widths for atmospheric lines in the Sun  from the tables by Moore,
Minnaert \& Houtgast (1966); we then adjusted the spectral resolution and line
strengths until they matched the resolution of GIRAFFE spectra and the
intensity of telluric features at the moment of observations.

Our stellar spectra were then divided by the adjusted synthetic spectrum
of the telluric lines,
cleaning fairly well the [O I] line from telluric contaminations. Finally, a
coadded spectrum was obtained from these cleaned spectra;  the final S/N ratio
is always high (S/N $>$ 100, and up to 300, depending on the stellar magnitude
and centering of the star on the fibre), as estimated from the averages
computed in several small intervals free of lines along the spectra.  

The blaze function was removed with standard IRAF tasks. After this, a refined
continuum tracement was derived as follows.  First, we summed (for each
grating) the spectra of all cluster stars; then we selected  a fair number of
fiducial regions of continuum in the resulting master spectrum (of very high
S/N ratio). By using a specialized set of commands in the ROSA  spectrum
analysis package (Gratton 1988), the final continuum placement was done   using
these fiducial points. 

Equivalent widths\footnote{$EW$s are available upon request from the first author} 
($EW$) were measured as described in detail in Bragaglia et
al. (2001); in particular, in the iterative clipping to derive a local
continuum around each line, after several checks we decided that a clipping 
factor of 2 for stars cooler than 4600~K and a factor 1 for warmer stars was
the optimal choice for NGC 2808.

\subsection{Iron abundances}

We started from the line list described in Gratton et al. (2003) and
extensively used in the analysis of high resolution spectra of GC stars (see 
Carretta et al. 2004a,b; Gratton et al. 2001). Atomic parameters  for the
subset of lines falling in the spectral range covered by gratings  HR11 and
HR13 are those given in that paper, as well as the reference solar abundances
used (computed using the Kurucz 1995 model atmospheres grid, see below).

However, the original line list was optimized for a higher resolution than the
present one, hence we had to cull out a few Fe I lines whose abundances were
systematically discrepant, likely because of blends.
We ended up with typically 15 to 18 Fe I lines safely measurable in stars
with only HR11 observations, 18 to 22 lines with only HR13
spectra, and up to 30 or 40 lines for stars observed with both gratings.
The number of measured Fe II 
lines ranges 
from zero to a maximum of 4.

Values of the microturbulence velocity $v_t$ were obtained by eliminating trends
of the abundances from Fe I lines with $expected$ line strength (see Magain
1984). We checked that the optimization for individual stars resulted in a much
smaller scatter in the derived abundances than using a mean value of $v_t$ as a
function of T$_{\rm eff}$ or $\log g$. For stars with observations only in
HR11  the uncertainty attached to $v_t$ is obviously larger, since only a few
lines  could be used. 

Final metallicities are obtained by choosing in the Kurucz (1995) grid of model
atmospheres (with the option for overshooting on) the model with the proper
atmospheric parameters whose abundance matches that derived from Fe I lines.
Average abundances of iron for NGC 2808 are [Fe/H]{\sc i} $=-1.10$ (rms=0.065) 
dex, from 123 stars and [Fe/H]{\sc ii} $=-1.16$ (rms=0.093) dex, from 90
objects. We do not think this difference is really relevant, since
abundances for Fe {\sc ii} rely on average on only two lines. 

Our metallicity, based on the analysis of medium resolution GIRAFFE spectra of
a large sample of RGB stars in NGC 2808, is in very good agreement with previous
results by Carretta et al. (2004a), who derived [Fe/H]{\sc i} $=-1.14$  
(rms=0.06) dex and [Fe/H]{\sc ii} $=-1.14$ (rms=0.13) dex from a sample of 20
red giants with UVES Red Arm spectra, covering a much wider spectral range.

\begin{figure}
\begin{center}
\includegraphics[bb=120 210 430 700, clip, scale=0.8]{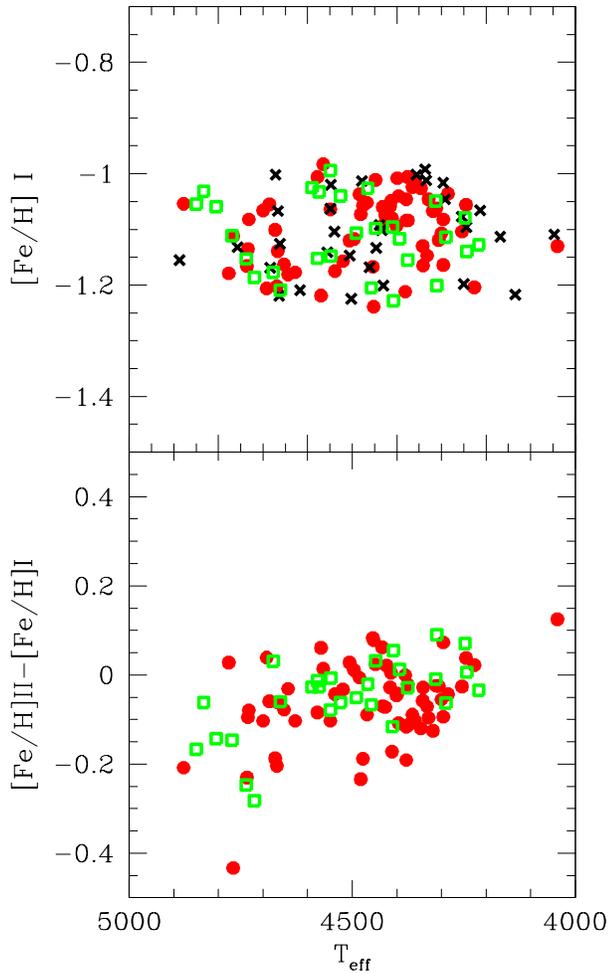}
\caption{Run of [Fe/H] ratio and of the Iron ionization equilibrium as a
function of temperatures for program stars in NGC 2808. Symbols and color
coding refer to the setup used: (red) filled circles indicate stars with 
both HR11 and HR13 observations,
(black) crosses for HR11 only, and (green) empty squares for HR13 only.}
\label{f:feteff}
\end{center}
\end{figure}

The distribution of resulting [Fe/H] values as a function of the temperatures is
shown in Figure~\ref{f:feteff}, with stars coded according to the grating they
were observed with. The scatter of the metallicity distribution is discussed in
Sect. 5.

\begin{table*}
\begin{center}
\caption[]{Adopted atmospheric parameters and derived iron abundances of RGB
stars in NGC 2808; nr indicates the number of lines used in the analysis.
The complete Table is available in electronic form.}
\begin{tabular}{rccccccrccc}
\hline
Star   &  T$_{\rm eff}$ & $\log$ $g$ & [A/H]  &$v_t$         & nr & [Fe/H]{\sc i} & $rms$ & nr &[Fe/H]{\sc ii} & $rms$ \\
       &     (K)        &  (dex)     & (dex)   &(km s$^{-1}$) &    & (dex)         &       &    & (dex)         &       \\
\hline
07183 &  4423  & 1.51 & $-$1.08& 1.66 & 41 &$-$1.08&   0.13   & 3 & $-$1.06&   0.17 \\
07315 &  4452  & 1.45 & $-$1.24& 1.54 & 33 &$-$1.24&   0.10   & 3 & $-$1.16&   0.17 \\
07536 &  4247  & 1.18 & $-$1.09& 1.45 & 23 &$-$1.08&   0.13   & 3 & $-$1.01&   0.11 \\
07558 &  4692  & 1.87 & $-$1.22& 1.30 & 33 &$-$1.21&   0.12   & 3 & $-$1.17&   0.18 \\
07788 &  4461  & 1.53 & $-$1.17& 1.40 & 18 &$-$1.17&   0.14   &   &	   &	    \\
08198 &  4617  & 1.81 & $-$1.21& 1.21 & 13 &$-$1.21&   0.14   &   &	   &	    \\
08204 &  4467  & 1.63 & $-$1.05& 1.35 & 38 &$-$1.05&   0.15   & 3 & $-$1.14&   0.14 \\
08603 &  4292  & 1.24 & $-$1.05& 1.65 & 19 &$-$1.05&   0.12   &   &	   &	    \\
08679 &  4734  & 1.72 & $-$0.25& 1.38 & 21 &	   &   0.20   & 3 & $-$0.79&   0.07 \\
08739 &  4213  & 1.13 & $-$1.07& 1.82 & 19 &$-$1.07&   0.12   &   &	   &	    \\
08826 &  4565  & 1.66 & $-$0.99& 1.22 & 39 &$-$0.98&   0.15   & 3 & $-$0.97&   0.08 \\
\hline
\end{tabular}
\label{t:atmpar}
\end{center}
\end{table*}

\subsection{Sodium and oxygen abundances}

Abundances of O and Na rest on measured $EW$s. For Na, one or both the doublets
at 5672-88~\AA\ and at 6154-60~\AA\ (depending on the setup observed) are always
available. Derived average Na abundances were corrected for effects of
departures from the LTE assumption using the prescriptions by Gratton et al.
(1999). 

Oxygen abundances are obtained from the forbidden [O {\sc i}] lines at 6300 and
6363~\AA. 
The O lines were carefully inspected by eye; in some cases they
were measured interactively if the automatic measurement failed (e.g., because
of some residual asymmetries due to imperfect removal of telluric lines). In
this check we were also able to derive fairly reasonable upper  limits to the
$EW$s in a few stars.

The contribution to the forbidden [O {\sc i}] line from the Ni blend at
6300.34~\AA\ is not a source of concern: Carretta et al. (2004a) estimated that
the $EW$s of the [O {\sc i}] 6300.31~\AA\ line in RGB stars are
hardly affected by more than $\sim 0.5$ m\AA ~in NGC 2808.
Also, CO formation is not expected to
lead to significant corrections to the O abundances, given the rather high
temperatures of the stars and the low expected C abundances.

\section{Errors in the atmospheric parameters}

Errors in the derived abundances are affected by three main contributions
(errors in temperatures, in microturbulence velocities and in the
measurements of $EW$s), and by two less severe error sources (errors in
surface gravities and in the adopted model metallicity).
In the following, we will concentrate on the major error sources.

\paragraph{Errors in temperatures.} 
Bedin et al. (2000) estimated as 0.02 mag (one $\sigma$)  the effect of
differential reddening across the cluster area where our targets were
selected.  Using the calibrating relations by Alonso et al. (1999), the effect
of an error of 0.02 mag in $E(B-V)$  translates into an error of 41 K in
T$_{\rm eff}$. On the other hand, the photometric error in the adopted $V-K$
colors is given by the quadratic sum of errors in $V$ (estimated in a few
thousandths of mag by Bedin et al.) and in $K$ ($\sim 0.02$ mag, from 2MASS).
Since the two photometries (optical and IR) are independent there is no color
term and we adopt a photometric error of 0.02 mag in $V-K$. When summed in
quadrature with the error due to the differential reddening ($\sim 0.05$ mag in
$E(V-K)$), we obtain 0.054 mag. The adopted internal error in T$_{\rm eff}$ is
thus 44 K.

\paragraph{Errors in microturbulence velocities.}
We computed the quadratic mean of the 1 $\sigma$ errors in the slope of
the abundance-expected line strength relation from all stars. Afterward, we
used star 7183 and repeated the analysis changing $v_t$  until the 1$\sigma$
value from the original slope of the relation between line strengths and
abundances was reached. A simple comparison allows us to give an estimate of
1$\sigma$ error associated to $v_t$, which is 0.09 km s$^{-1}$. This error
is mainly random; systematics due to blending and to the continuum tracement
are  negligible, and a systematic contribution from errors in the $gf$s is not
likely, at least for iron lines.

\paragraph{Errors in measurement of equivalent widths.}
In order to estimate this contribution, we selected a subset of 63 stars with 
more than 25 measured Fe lines. The average rms scatter (0.131 dex) in Fe
abundance for these stars, divided by the square root of the typical average 
number of measured line (38), provides a typical internal error of 0.022 dex. 

\paragraph{}

Table~\ref{t:sensitivity} shows the sensitivity of the derived abundances to
variations in the adopted atmospheric parameters for Fe, Na and O; this is
obtained by re-iterating the analysis while varying each time only one of the
parameters of the amount shown in the Table.
This exercise was done for all stars in the sample, and the average value
of the slope corresponding to the average temperature ($\sim 4500$ K) in the
sample was used to estimate the internal errors in abundances. 
For iron, these amount to $\sim 0.05$ dex and 0.027 dex, due to the quoted 
uncertainties in T$_{\rm eff}$ and $v_t$.

The impact of errors in $EW$s is evaluated in 
Col. 7, where the average error from a single line is weighted by the square
root of the mean number of lines, given in Col. 6. This is done for iron and
for the other elements measured in this paper.

\begin{table*}
\begin{center}
\caption[]{Sensitivities of abundance ratios to variations in the atmospheric
parameters and to errors in the equivalent widths, as computed for a typical 
program star with T$_{\rm eff} = 4500$ K. 
The total error is computed as the quadratic sum
of the three dominant sources of error,
T$_{\rm eff}$, $v_t$ and errors in the $EW$s (Col. 8: tot.1) or as the sum
of all contributions (Col. 9: tot.2)}
\begin{tabular}{lrrrrrrrr}
\hline
\\
Ratio    & $\Delta T_{eff}$ & $\Delta$ $\log g$ & $\Delta$ [A/H] & $\Delta v_t$
&$<N_{lines}>$& $\Delta$ EW & tot.1 & tot.2\\
         & (+50 K)    & (+0.2 dex)      & (+0.10 dex)      & (+0.20 km/s) & & & (dex)& (dex)  \\
 (1) & (2) & (3) & (4) & (5) & (6) & (7) & (8) & (9) \\
\\
\hline
$[$Fe/H$]${\sc  i}  &  +0.047 &   +0.001 &$-$0.001 & $-$0.029 & 38 &+0.021  &0.059 &0.059  \\
$[$Fe/H$]${\sc ii}  &$-$0.040 &   +0.010 &  +0.017 & $-$0.011 &  3 &+0.076  &0.087 &0.089  \\
\hline
$[$O/Fe$]${\sc  i}  &  +0.046 & $-$0.001 &  +0.004 &   +0.009 &  2 &+0.093  &0.104 &0.104  \\
$[$Na/Fe$]${\sc ii} &$-$0.006 & $-$0.006 &$-$0.020 &   +0.005 &  4 &+0.066  &0.066 &0.070  \\
\hline
\end{tabular}
\label{t:sensitivity}
\end{center}
\end{table*}

Total errors, computed using only the dominant terms and including all the
contributions, are reported in Table~\ref{t:sensitivity}, in Cols. 8 and 9
respectively.

\section{Cosmic scatter and reddening in NGC 2808}

We can now evaluate the expected scatter in [Fe/H] due to the uncertainties in
T$_{\rm eff}$, $v_t$ and errors in $EW$s, and from Table~\ref{t:sensitivity} we
derive $\sigma_{\rm FeI}$(exp.)=$0.059 \pm 0.008$ dex (statistical error). The
inclusion of contributions due to uncertainties in surface gravity or model
metallicity does not alter our conclusions. This result can be compared to
the {\it bona fide} observed scatter $\sigma_{\rm FeI}$(obs.)=$0.063 \pm 0.008$ 
dex (statistical error) estimated as the average rms scatter that we obtain
using the 63 stars in our sample with at least 25 measured iron lines.

From the quadratic difference between observed and expected scatter, we
can derive a formal value of 0.022 dex for the intrinsic spread in metallicity
in NGC 2808. Taking into account the statistical errors attached, we could 
set a limit of $\lsim 0.05$ dex as the maximum spread in iron abundance allowed
in this cluster. Note that a $\sim 2 \sigma$ evidence of a spread of $\sim
0.02$ dex seems to be present between the average abundances of Na-poor and
Na-rich stars, in NGC 2808 (see below).

In summary,  we believe that the dominant contribution to the spread comes from
differential reddening: we are confident that the intrinsic spread is small,
though perhaps not negligible. The stellar population in NGC
2808 can be considered reasonably homogeneous in Fe content, within a few
hundredths of dex.

\section{Results and discussion: the Na-O anticorrelation}

Abundances of O and Na are listed in Table~\ref{t:abunao} (only available in
electronic form) together with the number of measured lines and the rms value
obtained for each species.

\begin{table*}
\begin{center}
\caption[]{Abundances of O and Na in NGC 2808. [Na/Fe] values are corrected for
departures from LTE. HR is a flag for the grating used 
(1=HR11 only, 2=HR11 and HR13, 3=HR13 only). Lim is a flag discriminating
between real detections and
upper limits in the O measurements (0=upper limit, 1=detection). The complete
Table is available electronically.}
\begin{tabular}{rrccrcccc}
\hline
Star     &   nr &  [O/Fe] & rms   &   nr &   [Na/Fe]&  rms   & HR & lim \\
\hline
07183  &    2 &  +0.109 & 0.285 &    4 &   +0.217 &  0.093 &  2 &  1 \\
07315  &    2 &  +0.384 & 0.041 &    4 & $-$0.132 &  0.069 &  2 &  1 \\
07536  &    2 &  +0.386 & 0.049 &    2 &   +0.004 &  0.067 &  1 &  1 \\
07558  &    2 &  +0.428 & 0.205 &    2 &   +0.041 &  0.063 &  2 &  1 \\
07788  &      & 	&	&    2 &   +0.503 &  0.102 &  3 &  1 \\
08198  &      & 	&	&    2 &   +0.126 &  0.029 &  3 &  1 \\
08204  &    2 &  +0.324 & 0.004 &    4 & $-$0.010 &  0.162 &  2 &  1 \\
08603  &      & 	&	&    2 &   +0.052 &  0.143 &  3 &  1 \\
08679  &    2 &$-$0.136 & 0.028 &    2 &   +0.303 &  0.036 &  1 &  1 \\
08739  &      & 	&	&    2 &   +0.101 &  0.053 &  3 &  1 \\
08826  &    1 &$-$0.605 &	&    3 &   +0.470 &  0.167 &  2 &  1 \\
09230  &      & 	&	&    2 &   +0.659 &  0.081 &  3 &  1 \\
09724  &    2 &$-$0.171 & 0.091 &    4 &   +0.426 &  0.181 &  2 &  1 \\
09785  &      & 	&	&    2 &   +0.471 &  0.027 &  1 &  1 \\
10012  &    2 &$-$0.235 & 0.164 &    4 &   +0.369 &  0.151 &  2 &  1 \\
\hline
\end{tabular}
\label{t:abunao}
\end{center}
\end{table*}

\begin{figure}
\begin{center}
\includegraphics[bb=10 140 560 680, clip, scale=0.5]{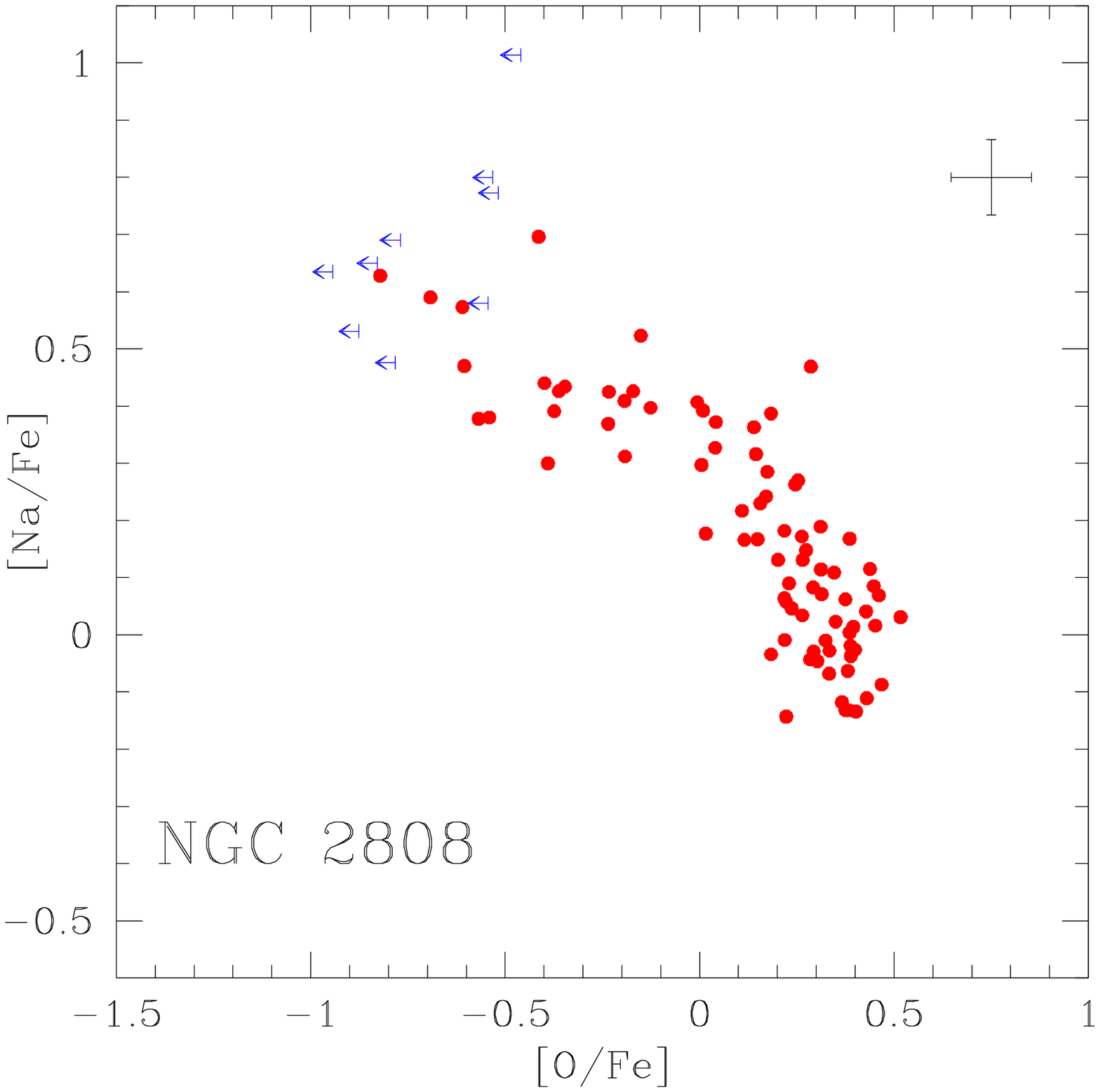}
\caption{[Na/Fe] ratio as a function of [O/Fe] for red giant stars in NGC 2808.
Na abundances do include the corrections for departures from LTE following
Gratton et al. (1999). Upper limits in [O/Fe] for a few stars are indicated as
blue arrows. The error bars take into account the uncertainties in
atmospheric parameters and $EW$s.}
\label{f:antim28}
\end{center}
\end{figure}

The [Na/Fe] ratio as a function of [O/Fe] ratio is displayed in
Figure~\ref{f:antim28} for each of the red giant stars with both O and Na
detections in NGC 2808, and for a few stars in which only upper limits in the
$EW$s of the [O I] 6300~\AA\ line were measured. The high resolution and high
S/N ratio of our spectra allow us to confidently reach stars down to 
[O/Fe]$\sim -1$.

The classical Na-O anticorrelation is clearly present also in this cluster, as
already shown by Carretta et al. (2004a): stars sharing the same position along
the giant branch show very different O and Na content. 

\begin{figure}
\begin{center}
\includegraphics[bb=130 150 440 690, clip, scale=0.8]{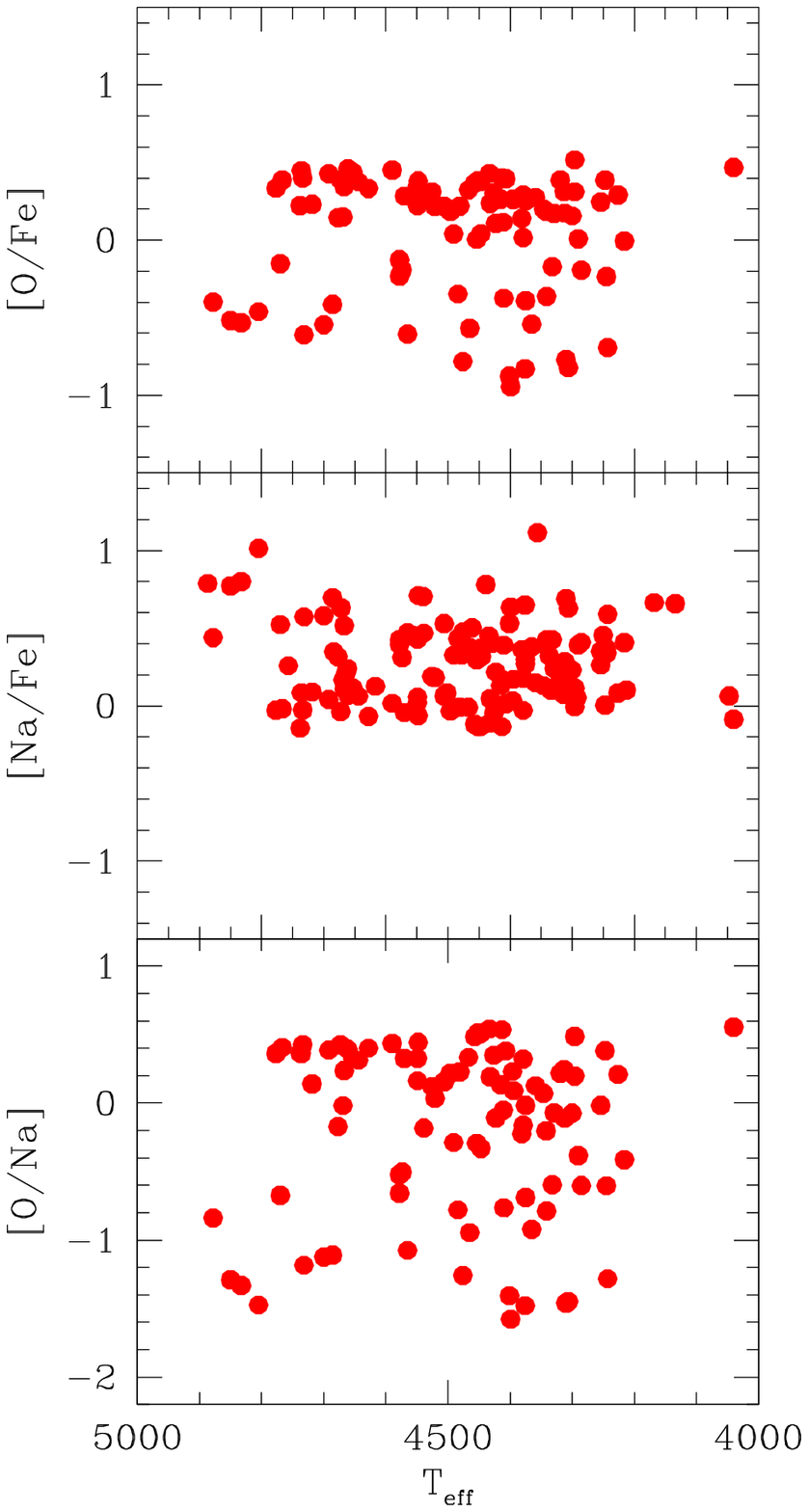}
\caption{Run of [O/Fe] (upper panel), [Na/Fe] (middle panel) and [O/Na] ratios
(lower panel) as a function of the evolutionary status (as represented by the 
effective temperature) for stars in our sample in NGC 2808.}
\label{f:naoteff}
\end{center}
\end{figure}

There is scarce -if at all- evidence of an internal origin  in the very same
stars that we presently observe for this phenomenon. There is no
evidence that O-poor, Na-rich stars are segregated in particular regions of the
sampled  RGB (upper and middle panels in Figure~\ref{f:naoteff}). 
This is at odds with expectations if evolutionary processes, due to some
extra-mixing, are at work as the stars climb along the RGB, bringing to the
surface more and more  material processed by nuclear proton-capture reactions.
The theoretical scenario would imply a bunch of heavily altered Na-rich,
O-depleted stars at the bright (cool) end of the RGB. This is not seen,
confirming early results by Carretta et al. (2003) in this cluster: the spread
in abundances seen at every luminosity along the RGB tells us that, whatever
the mechanism producing  the alterations is, these anomalies are likely to be
established well before the stars begun to move towards more advanced
evolutionary phases (see also Gratton et al. 2001 and Carretta et al. 2004b for
studies in scarcely evolved cluster stars).

Since Na or O saturates at the edges of the distribution function 
(the [O/Fe] ratio, for instance, levels off to the average value typical of
halo field stars, whereas [Na/Fe] still varies), the ratio
O/Na appears the best indicator to trace the stars distribution along the Na-O
anticorrelation because this ratio does continue to vary even at extreme
values. Excluding the non-members, we have 82 stars where O is detected and 9
other  with a robust upper limit.
However, we have Na abundances available for all the 123 member stars examined,
thus we are able to "project'' every star with no direct O determination along
the locus defined by the global Na-O anticorrelation, whose shape is very well
established.

\begin{table*}
\begin{center}
\caption[]{References for the [O/Fe] and [Na/Fe] ratios from high resolution
analyses in globular clusters used to derive the overall Na-O anticorrelation
shape.}
\begin{tabular}{lll}
\hline
Cluster  &  stars &  Reference\\
\hline
NGC 104 (47 Tuc)  &  SGB+TO &  Carretta et al. (2004b)    \\
                  &  RGB    &  Norris \& Da Costa (1995) \\ 
                  &  RGB    &  Carretta (1994)           \\ 
NGC 288           &  RGB    &  Shetrone \& Keane (2000)  \\ 
NGC 362           &  RGB    &  Shetrone \& Keane (2000)  \\ 
NGC 2808          &  RGB    &  Carretta et al. (2004a)    \\ 
                  &  RGB    &  present study             \\ 
NGC 3201          &  RGB    &  Gonzalez \& Wallerstein (1998) \\ 
NGC 5272 (M 3)    &  RGB    &  Cohen \& Melendez (2005)  \\ 
                  &  RGB    &  Sneden et al. (2004)      \\ 
NGC 5904 (M 5)    &  RGB    &  Ivans et al. (2001)       \\ 
NGC 6121 (M 4)    &  RGB    &  Ivans et al. (1999)       \\ 
NGC 6205 (M 13)   &RGB+SGB+TO&  Cohen \& Melendez (2005)  \\ 
                  &  RGB    &  Sneden et al. (2004)      \\ 
NGC 6254 (M 10)   &  RGB    &  Kraft et al. (1995)       \\ 	     
NGC 6397          &  SGB+TO &  Carretta et al. (2005)+Gratton et al. (2001) \\ 
                  &  RGB    &  Norris \& Da Costa (1995) \\ 
                  &  RGB    &  Carretta (1994)           \\ 
NGC 6528          &  RHB    &  Carretta et al. (2001)    \\ 
NGC 6715 (M 54)   &  RGB    &  Brown et al. (1999)       \\ 		     
NGC 6752          &  SGB+TO &  Carretta et al. (2005)+Gratton et al. (2001) \\ 
                  &  RGB    &  Yong et al. (2003)        \\ 
                  &  RGB    &  Norris \& Da Costa (1995) \\ 
                  &  RGB    &  Carretta (1994)           \\ 
NGC 6838          &RGB+SGB+TO& Ramirez \& Cohen (2002)   \\ 		     
NGC 7006          &  RGB    &  Kraft et al. (1998)       \\ 
NGC 7078 (M 15)   &  RGB    &  Sneden et al. (1997)      \\ 		     
Pal 5             &  RGB    &  Smith et al. (2002)       \\  
Pal 12            &  RGB    &  Cohen (2004)              \\ 		     
Ter 7             &  RGB    &  Tautvaisiene et al. (2004)\\  
\hline
\end{tabular}
\label{t:refertab}
\end{center}
\end{table*}

\begin{figure}
\begin{center}
\includegraphics[bb=10 140 590 710, clip, scale=0.4]{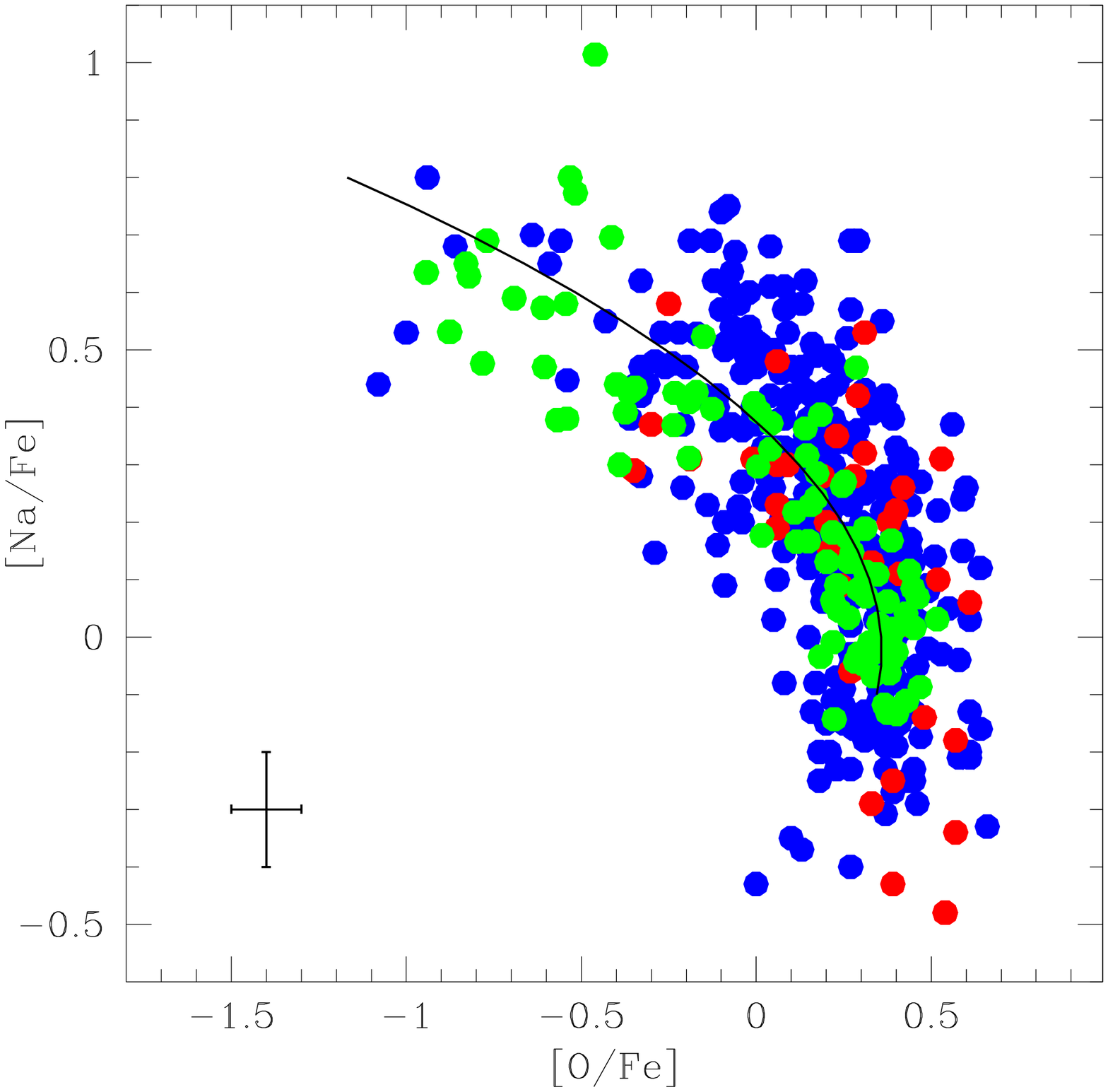}
\caption{Global Na-O anticorrelation (solid black line) superimposed to a
collection of stars in about 20 globular clusters. Blue points are RGB
stars from literature studies; red points are scarcely evolved stars (turnoff
or subgiant stars) from Gratton et al. (2001) and Carretta et al. (2004); 
green points are RGB stars in NGC 2808 from the present study.}
\label{f:relazione}
\end{center}
\end{figure}

This is shown in Figure~\ref{f:relazione} where the general shape of the Na-O
anticorrelation is drawn using a collection of literature 
data for almost
400 stars in about 20 globular clusters (47 Tuc, NGC 6752, NGC 6397, M 13, M 3,
M 5, NGC 3201, Ter 7, Pal 5, M 4, NGC 288, NGC 362, NGC 7006, M 15, M 10, Pal
12, M 71, NGC 6528, M 54; references and evolutionary status of the observed
stars are listed in Table~\ref{t:refertab}).
We took into account, whenever stated in the original papers, the
different adopted solar abundances, bringing them to our reference scale 
(Gratton et al. 2003). In this Figure most of the stars (blue points) are
evolved red giants, but the anticorrelation is well followed even by
scarcely evolved cluster stars (about 40 turnoff and subgiant stars, red 
points) as well as by RGB stars of the present study (green points).

Apart from providing us with an useful tool to exploit also stars having no 
direct O observations, this plot suggests some interesting points:
\begin{itemize}
\item[(i)] clusters of any metallicity in the range typical of these systems 
are represented in this plot, from M 92 ([Fe/H]$=-2.16$ dex, 
Carretta \& Gratton 1997) to NGC 6528 ([Fe/H]=$+0.07$ dex, Carretta et al. 2001);
\item[(ii)] a large range of physical properties (total mass, concentration, 
density) is sampled by these clusters; 
\item [(iii)] all kind of HB morphologies (from red clump only to very extended
blue HBs) are present among the clusters used to produce this plot;
\item[(iv)] all cluster stars, irrespective of their evolutionary status (from
scarcely evolved stars near the cluster turn-off to bright giants near the RGB
tip), seem to follow the same locus in this plane.
\end{itemize}

Moreover, some of the studied clusters (e.g. Pal 12, M 54, Ter 7) have been
very  likely originated in extragalactic objects, later accreted by our Galaxy.

From these evidences we can conclude safely enough that whatever the mechanism
responsible of the anticorrelation is, it must be an $intrinsic$ property of a
globular cluster, an universal feature of these objects.
Moreover, point (iv) supports the idea that it is likely related to 
the cluster formation process itself, since it is already in place among 
unevolved stars formed in the first few 10$^7$-$10^8$ years from the
beginning of the star formation in each cluster. 

In fact, unevolved low mass stars do not reach the high temperatures required
for activating proton capture chains as the NeNa and MgAl; moreover they do not
possess efficient convective envelopes. It follows that any variations in the
abundances of $p$-capture elements (O, Na, Mg, Al) must be likely already 
imprinted in the gas out of which these stars formed. The typical times
for the release of ejecta enriched in these elements (such as those from the
most favorite candidate polluters - intermediate-mass AGB stars - see
Gratton et al. 2004 and Carretta et al. 2005 for a detailed discussion) are
just a few 10$^7$-$10^8$ years, since these are among the first objects to
evolve and die in a globular clusters (see also D'Antona et al. 2005).

\begin{figure}
\begin{center}
\includegraphics[bb=10 140 590 710, clip, scale=0.4]{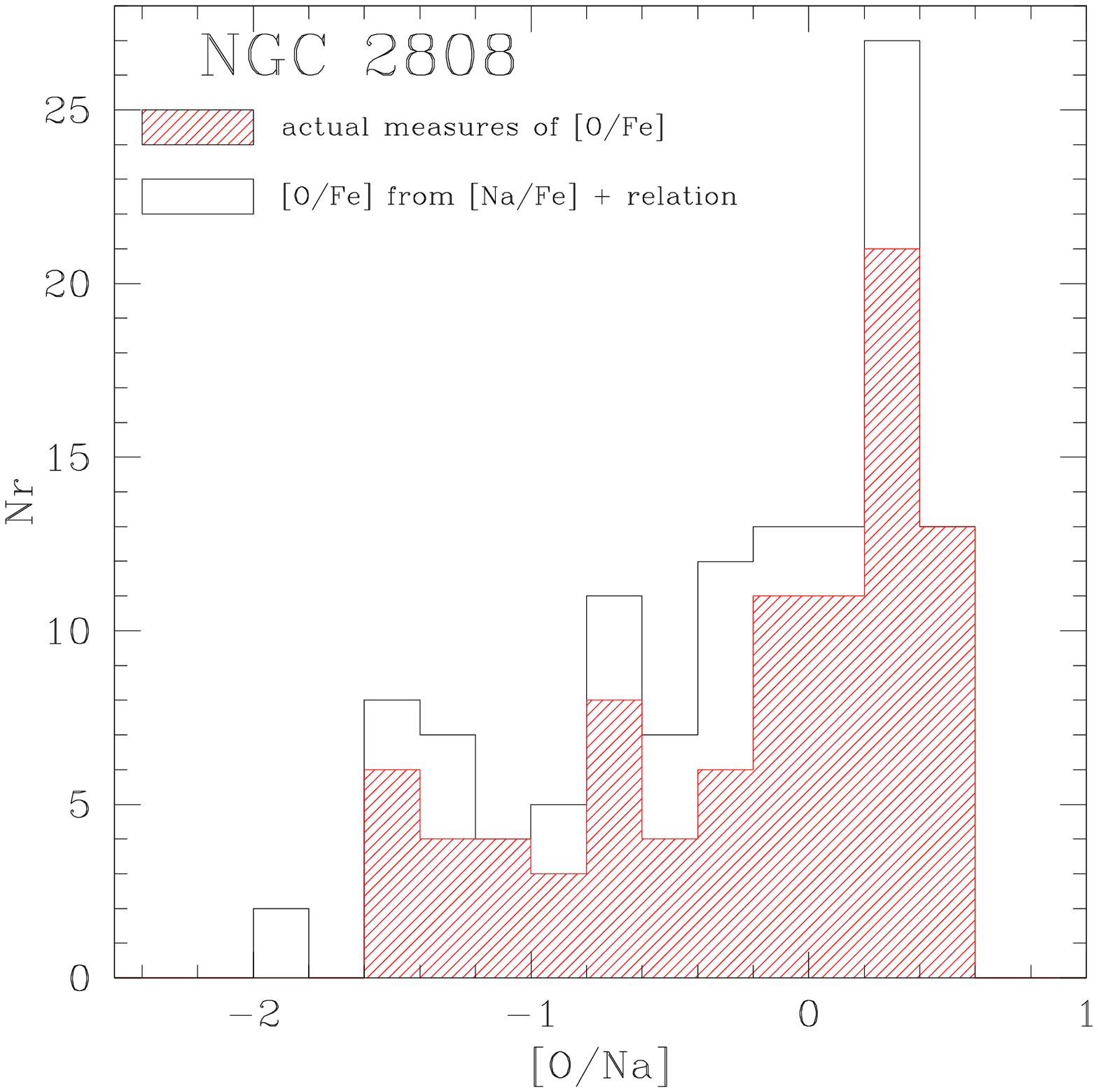}
\caption{Distribution function of the [O/Na] ratios along the Na-O
anticorrelation in NGC 2808. The dashed area is the frequency histogram
referred to actual detection of O in stars, whereas the empty histogram is
obtained by using the global anticorrelation relationship to obtain abundances
of O also for stars with no observation with HR13 and/or only upper limit in O
abundance.}
\label{f:histom28}
\end{center}
\end{figure}

The distribution function of stars along the Na-O anticorrelation in NGC 2808
is shown in Figure~\ref{f:histom28}, where the ratio [O/Na] from our data is
used. The dashed area shows the distribution obtained by using only actual
detections or carefully checked upper limits. The empty histogram is derived by
following the overall Na-O anticorrelation, in order to get [O/Fe] values even
for stars with no observations in HR13.

The bulk of stars along the RGB in NGC 2808 is peaked at [O/Fe] $\sim 0.28$ (and
[O/Na] $\sim 0.17$ dex) with
a scatter of 0.12 dex around this value; this should represent the ``normal" O
content of the ejecta by massive, type II supernovae (SNe), typical of
halo objects. We can identify this group of stars as the counterpart of the 
general halo field objects.

However, the distribution of stars continues down to very low
[O/Na] values in NGC 2808. 
A clearcut division into well defined groups is rather
hard to apply, since the appearance of the distribution is more similar to an 
extended tail, starting from normal-halo O values. However, if we take as a
working hypothesis [O/Na]$=-0.3$ and $-1.0$ as boundaries, we can tentatively
identify two other groups of O-poor and super O-poor stars, peaking at [O/Fe],
[O/Na]$= -0.21, -0.62$ and $-0.73, -1.40$ dex, respectively.

Although the statistical significance is admittedly not very high (the rms
scatter in each group is about 0.2 dex) the average Fe abundance is increasing,
going from [Fe/H]$=-1.113 \pm 0.008$ ($\sigma=0.067$ dex, 74 stars) in the group
of O-normal stars, to [Fe/H]$=-1.104 \pm 0.011$ ($\sigma=0.052$ dex, 27 stars),
and [Fe/H]$=-1.079 \pm 0.014$ ($\sigma=0.064$ dex, 21 stars) for the O-poor and
super O-poor groups.  
This is very interesting, since it is in qualitative agreement with what we
expect if O-depleted stars are also enriched in He from the likely same
polluting source (intermediate-mass AGB stars?). In fact, in stars with the
same original metal abundance, an increase in He abundance would be seen as an
increased strength of metallic lines (B\"ohm-Vitense 1979). 

This is exactly the trend that we observe in NGC 2808, even if more
significative conclusions on the intrinsic Fe spread are hampered by the
presence of a small differential reddening in this cluster.  However, this
effect of decreasing Fe cannot be explained by it, which would instead produce
fluctuations around a mean value.

$Viceversa$, we can reverse this line of thought and ask what mass fraction Y
of He may be associated to these three groups, simply starting from the
definition of the logarithmic ratio [Fe/H], assuming the number of H atoms
proportional to the mass fraction X and neglecting the contribution of heavy
species Z. We found that in order to reproduce the different [Fe/H] between the
O-normal group (likely having a primordial Y=0.24) and the super O-poor group 
we need to consider a value Y=0.30 for the latter. The intermediate group
should have a mass fraction Y of about 0.26.

Interestingly, we do not find any evidence of a group having a large He
enrichment, about Y=0.40: this would correspond to a mean value of
[Fe/H]$\sim -1.010$ dex, 0.1 dex more metal-rich than the dominant population.
This would exceed the intrinsic scatter carefully evaluated in Sect. 5.
If confirmed, this would put strong constraints on every model concerning star
formation, in one or multiple generations in a globular cluster. In particular,
the model proposed very recently by D'Antona et al. (2005) 
predicts about 20\% of stars being generated in a stage of star
formation from ejecta of massive AGB stars with Y=0.40. We would expect to
reveal such stars also along the RGB. However, stars very He-enriched would
have a turnoff mass lower than about 0.65 M$_\odot$ (D'Antona et al. 2005), 
and if the mass loss rate does not depend strongly on Y these stars 
could have left the RGB before reaching the He-flash point. In this case, these 
objects could not be really represented in our observed sample.

Notice also that the fractions we found for the different groups on the
RGB support this view.
From the numbers of stars in the RHB and BHB of NGC 2808 as reported in D'Antona \&
Caloi (2004) and from the histogram in their Fig. 1 we derived a ratio
1:0.65:0.38 for the RHB clump, the so-called EBT1 (the bulk of stars at
brighter magnitudes in the BHB) and the sum of stars in the EBT2+EBT3 groups
(the faint part of the BHB, see Bedin et al. 2000). As explained in the
modeling by D'Antona et al. (2005), these are the likely outcome of stars
starting their life with a He mass fraction Y=0.24, $\sim 0.27$ and 0.40,
respectively. 

On the other hand, following the above discussion, we can infer that  in our
sample along the RGB in NGC 2808 we found 74 O-normal stars, likely with a
(``cosmological") He content Y=0.24, and 27+21 stars in the O-poor and super
O-poor subgroups respectively, with Y increasing up to Y=0.30. In these case
the ratio is again 1:0.65 for the O-normal:(O-poor+super O-poor) populations,
supporting the view that we are sampling the progenitors of RHB stars and of
the bulk of brighter BHB stars.
The super He-rich progenitors of the extreme BHB stars could be simply missing
in our sample if they never completed their RGB evolution.

\begin{figure}
\begin{center}
\includegraphics[bb=50 190 390 690, clip, scale=0.6]{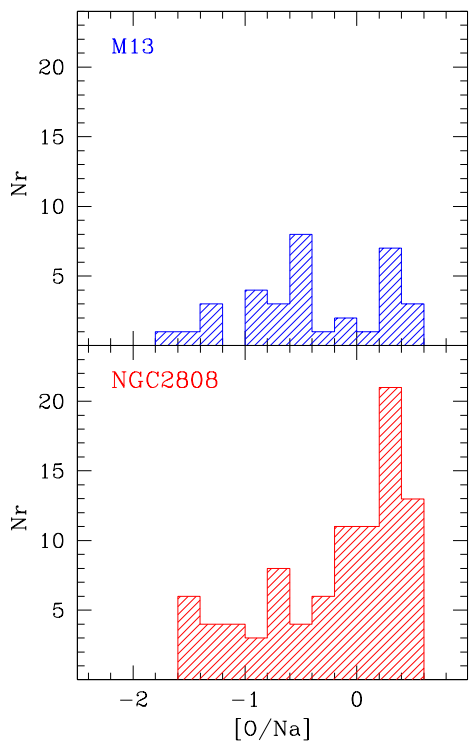}
\caption{Distribution function of the measured [O/Na] ratios along the Na-O
anticorrelation in M 13 (Sneden et al. 2004) and NGC 2808. O and Na abundances
in M 13 are shifted to our scale by correcting for different adopted solar
abundances; for M 13 we used the Na abundances corrected for departures from
LTE.}
\label{f:m13histo}
\end{center}
\end{figure}

In Figure~\ref{f:m13histo} we compare the observed distributions of RGB stars
along the global Na-O anticorrelation in NGC 2808 (present study) and M 13
(Sneden et al. 2004), the template cluster as far as the Na-O anticorrelation
is concerned. 
Abundances for Na and O in M 13 are corrected for the adoption of  solar
abundances different from ours; for Na we used the values corrected for effects
of departures from LTE using the same prescriptions by Gratton et al. (1999),
as we did for NGC 2808.
This (homogeneous) comparison shows that:
\begin{itemize}
\item in both clusters three groups of stars (O-normal, O-poor and super
O-poor) can be seen distributed along the global Na-O anticorrelation. In
particular, we note that M 13 is not anymore the only cluster in which stars
with very low O abundances can be observed. Although Sneden et al. (2004) do
not distinguish between limits and actual detections, stars with [O/Fe] ratios
as low as $\sim -1$ (within the uncertainties related to the analysis) are
found in both clusters;
\item whatever the mechanism responsible for the anticorrelation is, it produced
the same range in [O/Na] ratios;
\item the relative weights in the distribution functions along the Na-O
relation are different in the two clusters: a Kolmogorov-Smirnov test shows
only a $\sim 3$\% probability that both samples in Figure~\ref{f:m13histo}
are extracted from the same parent distribution;
\item at variance with M 13, most stars in NGC 2808 are O-normal. This might
be related to the HB morphologies, very different in the two clusters: NGC 2808
has a bimodal HB (HB parameter (B-R)/(B+V+R)=-0.49 from the on line
catalogue by Harris 1996) while the HB of M 13
(HB parameter = 0.97 from the same source) is populated only to the blue of
the RR Lyrae instability strip.
\end{itemize}

In NGC 2808 there seems to be a rather straightforward correspondence between
the stars along the RGB and their progeny on the HB: following the previous
discussion, it seems possible to identify RHB stars as those having
O-rich/Na-poor/He-poor composition when formed. The two other groups could
originate the BHB, for instance as in the scenario devised by D'Antona and 
coworkers.

However, the case of M 13 seems to show that more modeling is required, since
a relevant group of O-normal stars is present on the RGB, yet no RHB stars are
found in this cluster. Where have all the O-normal stars  gone, after the
He-flash, in M 13? We have to postpone a more thorough discussion to the
completion of the analysis of our whole sample of  clusters with different HB
morphologies.

\section{Summary and conclusions}

In this paper we have derived atmospheric parameters and elemental abundances
for about 120 red giant stars in the globular cluster NGC 2808. 

From the analysis of GIRAFFE spectra sampling the forbidden [O {\sc i}] lines \
and the
Na doublets at 5682-88 and 6154-60~\AA\ we measured Na and O abundances for a
large sample of stars. We also derived the distribution function of stars in
[O/Na], i.e. along the Na-O anticorrelation, the well known signature of
proton-capture reactions at high temperature.

We found that the bulk of stars along the RGB in NGC 2808 has normal O (and Na)
content, typical of field halo stars, where we are seeing 
predominantly the contributions of yields from massive type II supernovae.
However, we tentatively identified also two other groups of stars, whose O
content is depleted or even strongly depleted. 

Evidence of accompanying He-enrichment comes from the average Fe abundances in
these two groups, although the statistical significance is not very high.
However, the qualitative agreement between the observations and the theoretical
prediction for pollution by He-rich and O-depleted matter provided by a
previous generation of IM-AGB stars supports a scenario where a fraction of the
presently observed stars was formed from polluted or heavily polluted
intracluster gas.

In this regard, the distribution function of stars in [O/Na] for NGC 2808 is 
similar to the one seen in M 13, a template cluster as far as the 
the Na-O anticorrelation is concerned. 
This is interesting, since the HB morphology of these two cluster is very
different.

The average metallicity we found for NGC 2808 is [Fe/H]$=-1.10$
(rms=0.065 dex, from 123 stars). We also found some evidence of a small 
intrinsic spread in metallicity, but more definitive conclusions are hampered
by the presence of a small differential reddening.

\begin{acknowledgements}
{This
publication makes use of data products from the Two Micron All Sky Survey,
which is a joint project of the University of Massachusetts and the Infrared
Processing and Analysis Center/California Institute of Technology, funded by
the National Aeronautics and Space Administration and the National Science
Foundation. This work was partially funded by Cofin 2003029437 (P.I. Raffaele
Gratton) "Continuit\`a e discontinuit\`a
nella formazione della nostra Galassia" by Ministero Universit\`a e Ricerca
Scientifica, Italy. }
\end{acknowledgements}

\end{document}